%% file: ci_rs.tex
\documentclass[conference]{IEEEtran}
\IEEEoverridecommandlockouts
\usepackage{cite}
\usepackage{amsmath,amssymb,amsfonts}
\usepackage{algorithmic}
\usepackage{algorithm}
\usepackage{graphicx}
\usepackage{makecell}
\usepackage{textcomp}
\usepackage{arydshln}
\usepackage{xcolor}
\def\BibTeX{{\rm B\kern-.05em{\sc i\kern-.025em b}\kern-.08em
    T\kern-.1667em\lower.7ex\hbox{E}\kern-.125emX}}

\newcommand{\NEAT}{\textbf{NEAT} }
\newcommand{\NEATe}{\textbf{NEAT}}

\begin{document}

\title{\NEATe: A Label Noise-resistant Complementary Item Recommender System with Trustworthy Evaluation
}

\author{\IEEEauthorblockN{Luyi Ma, Jianpeng Xu, Jason H.D. Cho, Evren Korpeoglu, Sushant Kumar, Kannan Achan}
\IEEEauthorblockA{\textit{Walmart Global Tech},
Sunnyvale, CA \\
{\{luyi.ma, jianpeng.xu, jason.cho, EKorpeoglu, sushant.kumar, kannan.achan\}}@walmart.com}
}

\maketitle

\begin{abstract}
The \textit{complementary item recommender system} (CIRS) recommends the complementary items for a given query item. 
Existing CIRS models consider the item co-purchase signal as a proxy of the complementary relationship, due to the lack of human-curated labels from the huge transaction records. 
These methods represent items in a complementary embedding space and model the complementary relationship as a point estimation of the similarity between items vectors. 
However, co-purchased items are not necessarily complementary to each other. For example, customers may frequently purchase \textit{bananas} and \textit{bottle water} within the same transaction, but these two items are not complementary. 
Hence, using co-purchase signals directly as labels will aggravate the model performance.
On the other hand, model evaluation will not be trustworthy if the labels for evaluation are not reflecting the true complementary relatedness. 
To address the above challenges from noisy labeling of the co-purchase data, we model the co-purchases of two items as a Gaussian distribution, where the mean denotes the co-purchases from the complementary relatedness, and covariance denotes the co-purchases from the noise.  
To do so, we represent each item as a Gaussian embedding and parameterize the Gaussian distribution of co-purchases by the means and covariances from item Gaussian embedding. 
To reduce the impact of the noisy labels during evaluation, we propose an independence test-based method to generate a trustworthy label set with certain confidence. 
Our extensive experiments on both the publicly available dataset and the large-scale real-world dataset justify the effectiveness of our proposed model in complementary item recommendations compared with the state-of-the-art models.
\end{abstract}




\maketitle

\input{intro}

\input{relatedwork}

\input{model}

\input{labels}

\input{experiments}

\section{Conclusions} \label{sec:conclusion}
In this paper, we proposed a label noise-resistant complementary item recommendation model named \NEAT to address the label noise issue for complementary item recommendation when the co-purchase data are used as labels. \NEAT learns the item representations as Gaussian embeddings, and assumes the co-purchase data as a Gaussian distribution, where the mean is the co-purchases from the true complementary relation, and the variance is the co-purchases from the noise. In addition, we developed a trustworthy label generation method for model evaluation to alleviate the impact of noisy labels in evaluation step. We performed extensive experiments on two real-world datasets and the results show the effectiveness of the proposed method over state-of-the-art models. 

\bibliographystyle{IEEEtran}
\bibliography{ci_rs}

\end{document}

%% file: intro.tex
\section{Introduction}
Item recommendation tasks in e-commerce industry are essential for improving user experiences by recommending related items to a query item. 
Different types of recommender systems have been proposed to address use cases under various aspects of the \textit{relatedness}, such as substitutional items (SI) recommendation and complementary items (CI) recommendation \cite{DBLP:conf/cikm/LiuGDGGBY20}\cite{DBLP:conf/kdd/McAuleyPL15}\cite{DBLP:conf/wsdm/WangJRTY18}.
In economics, a complementary item is a type of items whose appeal increases with the popularity of its complement. 
Therefore, complementary items usually have higher chances to be purchased together to complete the same shopping goal. 
For example, \texttt{Shampoo} and \texttt{Conditioner} are complementary to each other in order to fulfill the needs of \textit{shower supplies}; similarly, \texttt{TV} and \texttt{TV Mount} are also complementary items for \textit{TV entertainment} purposes. 
While SI recommendations have been extensively studied in the past \cite{DBLP:conf/sigir/ChenYYH0020}\cite{DBLP:conf/ijcai/ZhangYWCCN19}, complementary item recommender systems (CIRS) become increasingly important as they 
provide the customers with the opportunities to explore and interact with items that are complementary with what they have been interested in, 
and hence complete the customers' shopping journey by suggesting purchasing those items together. 

Although the complementary relationship between items seems well-defined, it is impossible to gain the ground truth of the complementary relationship for all item pairs from the catalogue. 
To mitigate the labeling challenge, a common practice is to indicate the complementary relationship using the co-purchase signal of two items  \cite{DBLP:conf/cikm/LiuGDGGBY20}\cite{DBLP:conf/kdd/McAuleyPL15}\cite{DBLP:conf/cikm/WanWLBM18}\cite{DBLP:conf/wsdm/WangJRTY18}.
These CIRS models usually represent each item as an embedding vector under co-purchase space, and the similarity between the items in the latent space reflects the frequency of co-purchases, and hence the complementary relatedness under their assumptions. 

However, co-purchased items are not necessarily complementary to each other. 
For example, certain popular items can appear in many transactions and hence be co-purchased frequently with items that are not complementary. 
Simply removing these popular items from all recommendations will hurt the results for item pairs with real complementary relations and decrease the business metrics (e.g., Gross Merchandise Value) of the recommender systems. 
Recently, Hao et al. proposed to annotate the co-viewed but not co-purchased item pairs as the negative labels and consider the co-purchased but not co-viewed item pairs as positive labels for learning  \cite{DBLP:conf/cikm/HaoZLDFSW20}.
However, co-view data are noisy by themselves as well. Cleaning noisy labels with another noisy data source is not trustworthy in general. 
Identifying and cleaning co-purchased non-complementary items is not feasible due to the lack of ground truths. 
Hence, it is challenging to learn the real complementary relationships between items pairs and evaluate the recommendation results with the noisy labels. 

To address the noisy label issue during training, 
we assume that the co-purchases of items are composed by two components: (a) co-purchases motivated by the true complementary relationships, and (b) co-purchases from other motivations (say, the noise). We directly model component (a) by the similarities or distances of item embeddings under the complementary space, and component (b) by the variance around (a). 
Hence, the co-purchase data can be assumed as a Gaussian distribution, where the mean is the co-purchases from the true complementary, and the variance is the co-purchases from the noise. 
To achieve this, instead of representing items as item embeddings under point estimation, we employ Gaussian embeddings \cite{DBLP:journals/corr/VilnisM14} with a mean vector and a covariance matrix to as item representations. 
The Gaussian distribution of the co-purchase data can be naturally parameterized by the item Gaussian embeddings and fit into the noisy co-purchase data by optimizing the \textit{expected likelihood} \cite{DBLP:journals/jmlr/JebaraKH04} between Gaussian embeddings. 
To this end, works such as \cite{DBLP:conf/cikm/LiuGDGGBY20}\cite{DBLP:conf/kdd/McAuleyPL15}\cite{DBLP:conf/cikm/WanWLBM18}\cite{DBLP:conf/wsdm/WangJRTY18} are special cases of this assumption which assume that all co-purchases occur under complementary relationships. 
They represent each item as a vector in the embedding space and the co-purchases or complementary relationships are calculated by the similarities between item embeddings.

To address the noisy label issue during evaluation, we follow the definition of complementary items and develop an independence test-based method to surface the item pairs with more complementarity as positive labels for evaluation.
Given a pair of co-purchased items, we treat the purchase of the individual item as a binary random variable and study the difference between observed co-purchase frequency and the expected co-purchase frequency under the independence assumption via Chi-squared independence test \cite{pearson1900x}.
Based on the definition of complementary items in economics, the purchases of them should be dependent and the observed co-purchase frequency should be larger then the expected independent co-purchase frequency due to the synergy effect between complementary items.
A set of co-purchase labels could be generated for evaluation by providing a predefined p-value, which controls the certainty of the label selection from the noisy observation. 
Although it is promising to use the selected label as the ground truth labels for training as well, the coverage of this set over the item catalogue is very limited and hence not feasible to be generalized for training purpose. 


In summary, we developed a label \textit{\textbf{N}}oise-r\textit{\textbf{E}}sist\textit{\textbf{A}}n\textit{\textbf{T}} CIRS model named \NEATe, which learns the complementary relationship by Gaussian embedding representation. 
In order to accurately evaluate the model performance, we created a trustworthy label set with controllable confidence via an independence test. 
Extensive experiments are conducted on the publicly available \textit{Instacart} dataset and a real-world large-scale dataset collected from www.walmart.com. The results demonstrated the effectiveness of \NEAT in modeling complementary relationships from co-purchase data, and the superior performance over the state-of-the-art models in CIRS.

The rest of the paper is structured in the following: 
the related work on CIRS is discussed in Section \ref{sec:related_work}. 
Section \ref{sec:FACIRS} describes the details of the proposed method \NEAT and Section \ref{label_generation} presents the trustworthy label creation for evaluation.
Experiment settings and results are reported in Section \ref{sec:experiments}. In the end, we conclude the paper in Section \ref{sec:conclusion}.

%% file: relatedwork.tex
\section{Related Work} \label{sec:related_work}
\subsection{Embedding-based Complementary Item Recommendations}
Embedding-based CIRS are most popular in recent work of CIRS. 
They treat each item as a vector in the embedding space and estimate the complementary relationship based on the distance between item vectors.
The first embedding-based method for CIRS was proposed in \cite{DBLP:conf/mlsp/BarkanK16}, which models the co-purchase of items by the similarity between the embeddings of the co-purchased items under the effective training paradigm of Skip-gram with Negative Sample (SGNS)  \cite{DBLP:conf/nips/MikolovSCCD13}.
Wan et. al. extended \cite{DBLP:conf/mlsp/BarkanK16} by incorporating user behavior into the modeling of the item-level complementary relationship with the user and item embeddings learned jointly \cite{DBLP:conf/cikm/WanWLBM18}. 
Besides of modeling the item embeddings with co-purchase data using SGNS, co-purchase data are also represented as item graphs in \cite{DBLP:conf/cikm/LiuGDGGBY20} \cite{DBLP:conf/kdd/McAuleyPL15} \cite{DBLP:conf/wsdm/WangJRTY18} and identifying the complementary relationships between items are treated as the link prediction tasks 
between item nodes.
They use the co-purchase records as labels for link predictions based on the distance between item embeddings.
To further improve the complementary recommendations, different types of auxiliary data are incorporated into the modeling.
Multi-modal data of items such as item descriptions and images are also included in \cite{DBLP:conf/recsys/ZhangLNC18} to learn the multi-modal representations of items.
The distance between vector embeddings of two co-purchased items in each modal's embedding space is minimized for complementarity measurement. 
Xu et al. in \cite{DBLP:journals/corr/abs-1911-12481} considers the last $l$ purchased items as the context to learn the attention-based encoder and represents the complementary relationship via the inner product between encoded item embeddings. This work is reduced to the \textbf{Item2Vec} model in \cite{DBLP:conf/mlsp/BarkanK16} without the context of the last $l$ items in the history.
Although the \textbf{P-companion} model in \cite{DBLP:conf/cikm/HaoZLDFSW20} pre-processes the co-purchase labels by removing co-view data and leverage the product-type information to improve the diversity, it still models the complementary relationship via the distance between item embeddings without addressing the noise in the data by parameters.
Despite of various auxiliary information such as graph structure, multi-modal data source, shopping context and product taxonomy, 
all these models are trying to build item embeddings in co-purchase space, and model the co-purchase data using the similarity or distance between the co-purchased items. Hence, these models will suffer from the noisy labels for learning complementary relations. 
In addition to the item-level complementary recommendation models, many in-basket recommendation models try to address the complete-the-basekt tasks.
For example, \textbf{BasConv} \cite{DBLP:conf/sdm/LiuWGAY20} leverages the heterogeneous graph embeddings to perform the in-basket recommendations; 
multiple intents in the same basket are modeled in \cite{DBLP:journals/corr/abs-2010-11419} to the in-basket recommendation.
Although these models capture the co-purchase pattern in the same basket to complete the basket, items in the same basket might not be complementary, for instance, a basket containing both grocery shopping and the household shopping. 
Even the items with the same shopping intent like grocery shopping might not be complementary. 
The goal of the in-basket recommendations focuses on completing the basket which cover various types of recommendations such as re-purchase, popular items and user preference in addition to co-purchases and complementary items. 
Hence, in-basket recommendation is out of the scope of the discussion in this paper. 

\subsection{Gaussian Embedding in Recommender Systems}
Gaussian embedding \cite{DBLP:journals/corr/VilnisM14} has been applied in recommender system in recent years, e.g., Gaussian embeddings for collaborative filtering  \cite{DBLP:conf/sigir/SantosPG17} and convolutional Gaussian embeddings for personalized item recommendations \cite{ijcai2019-367}.
They mainly use the Gaussian embeddings to address the different confidences of user/item representations introduced by the lack of user/item information or contradictions between user/item behaviors (e.g., item ratings and reviews by users).
However, these methods were not designed for CIRS and hence not applicable to address the unique challenges from CIRS. 

%% file: model.tex
\section{Label Noise-resistant Complementary Item Recommender Systems} \label{sec:FACIRS}

In this section, we first define the co-purchase records from transactions and then go through the details of modeling item-level complementary relationship as well as item representation for recommendations.

\subsection{Co-purchase Records from Transactions} \label{sample_co-purchase}
Let $v$ denote an item from the item set $\mathcal{V}$ and $b$ denote a transaction (a set of purchased items) from the transactions set $\mathcal{B}$ where $b = \{v_1, v_2, ...\}$.
A tuple $(v_i, v_j)$, $v_i \neq v_j$, from the same transaction $b$ can be considered as a pair of co-purchased items (i.e., a co-purchase record). 
To further distinguish the role of co-purchased items during training, inference and evaluation, we treat the first item in an item pair $(v_i, v_j)$ as the query item $q$ and the second item as the recommendation of $q$.

\subsection{Learning Item-level Complementary Relationship} \label{sec:item_level_complementary}
Learning the complementary relationship with the co-purchase data as labels could suffer from the label noisy, as co-purchased items are not necessarily complementary items.  
Previous CIRS models simply treat the co-purchased items as the positive label of complementary relationships and fit them by the distance between item embeddings in the embedding space.
Formally, given a pair of co-purchased items $(q, v)$, they try to maximize the density of a particular normal distribution at zero: $\mathcal{N}\left(0; \mbox{\boldmath$\mu$}_{q}-\mbox{\boldmath$\mu$}_{v},\mbox{\boldmath$\Sigma$}_{zero}\right) \propto - \left (\mbox{\boldmath$\mu$}_{q}-\mbox{\boldmath$\mu$}_{v} \right)^T \left ( \mbox{\boldmath$\mu$}_{q}-\mbox{\boldmath$\mu$}_{v}\right)$, where $\mbox{\boldmath$\Sigma$}_{zero}$ is zero and \mbox{\boldmath$\mu$} is the item embeddings, to bring the item embedding ($\mbox{\boldmath$\mu$}_{q}, \mbox{\boldmath$\mu$}_{v}$) closer in the embedding space.
Because these models do not consider the noise in the co-purchase labels, the distance between item embeddings is hardly reflecting the complementary relationship, even it might be a good approximation for co-purchases. 

To address the label noise issue for learning complementary relationship, as aforementioned, we model the co-purchase data as a Gaussian distribution, where the mean is the co-purchases from the true complementary, and the variance is the co-purchase from the noise. 
In order to do so, we consider each item $v \in \mathcal{V}$ as a Gaussian embedding 
$\mathcal{N}\left(x; \mbox{\boldmath$\mu$}_{v},\mbox{\boldmath$\Sigma$}_{v}\right)$, where $\mbox{\boldmath$\mu$}_{v} \in \mathbb{R}^d$ is the mean vector and $\mbox{\boldmath$\Sigma$}_{v} \in \mathbb{R}^{d\times d}$ is the covariacne matrix in the $d$-dimensional embedding space, which models the variation in the co-purchase behavior of $v$.
While the inner product between vectors of two items is used to model their complementary relationship from the co-purchase record in the literature \cite{DBLP:conf/cikm/WanWLBM18}\cite{DBLP:conf/mlsp/BarkanK16}, we compute  the \textit{expected likelihood} \cite{DBLP:journals/jmlr/JebaraKH04} as the inner product of two Gaussian embeddings \cite{DBLP:journals/corr/VilnisM14} to parameterize the Gaussian distribution of complementary relationships.
Given an item pair $(q, v)$, the \textit{expected likelihood} between their Gaussian embeddings is defined in Equation \ref{eq:expected_likelihood}, which is the probability density of a Gaussian distribution at zero, $\mathcal{N}\left(0; \mbox{\boldmath$\mu$}_{q}-\mbox{\boldmath$\mu$}_{v},\mbox{\boldmath$\Sigma$}_{q}+\mbox{\boldmath$\Sigma$}_{v}\right)$.

\begin{align} \label{eq:expected_likelihood}
    E(q, r) &= \int_{x\in \mathbb{R}^d} \mathcal{N}\left(x; \mbox{\boldmath$\mu$}_{q},\mbox{\boldmath$\Sigma$}_{q}\right) \mathcal{N}\left(x; \mbox{\boldmath$\mu$}_{v},\mbox{\boldmath$\Sigma$}_{v}\right) dx \nonumber \\
    &= \mathcal{N}\left(0; \mbox{\boldmath$\mu$}_{q}-\mbox{\boldmath$\mu$}_{v},\mbox{\boldmath$\Sigma$}_{q}+\mbox{\boldmath$\Sigma$}_{v}\right)
\end{align}

Hence, $\mathcal{N}\left(x; \mbox{\boldmath$\mu$}_{q}-\mbox{\boldmath$\mu$}_{v},\mbox{\boldmath$\Sigma$}_{q}+\mbox{\boldmath$\Sigma$}_{v}\right)$ denotes the Gaussian distribution of the co-purchase data between $(q, v)$, where the mean is the difference between two items mean vectors in complementary space and the covariance matrix combines the variance of each individual items. 
The probability density at zero, $\mathcal{N}\left(0; \mbox{\boldmath$\mu$}_{q}-\mbox{\boldmath$\mu$}_{v},\mbox{\boldmath$\Sigma$}_{q}+\mbox{\boldmath$\Sigma$}_{v}\right)$, represents the likelihood of observing a co-purchase record of $(q, v)$ when considering both their complementary relationship ($\mbox{\boldmath$\mu$}_{q}-\mbox{\boldmath$\mu$}_{v}$) and variations of purchase behaviors ($\mbox{\boldmath$\Sigma$}_{q}+\mbox{\boldmath$\Sigma$}_{v}$).

\begin{figure} [ht]
    \begin{minipage}[t]{0.48\textwidth}
        \includegraphics[width=\textwidth]{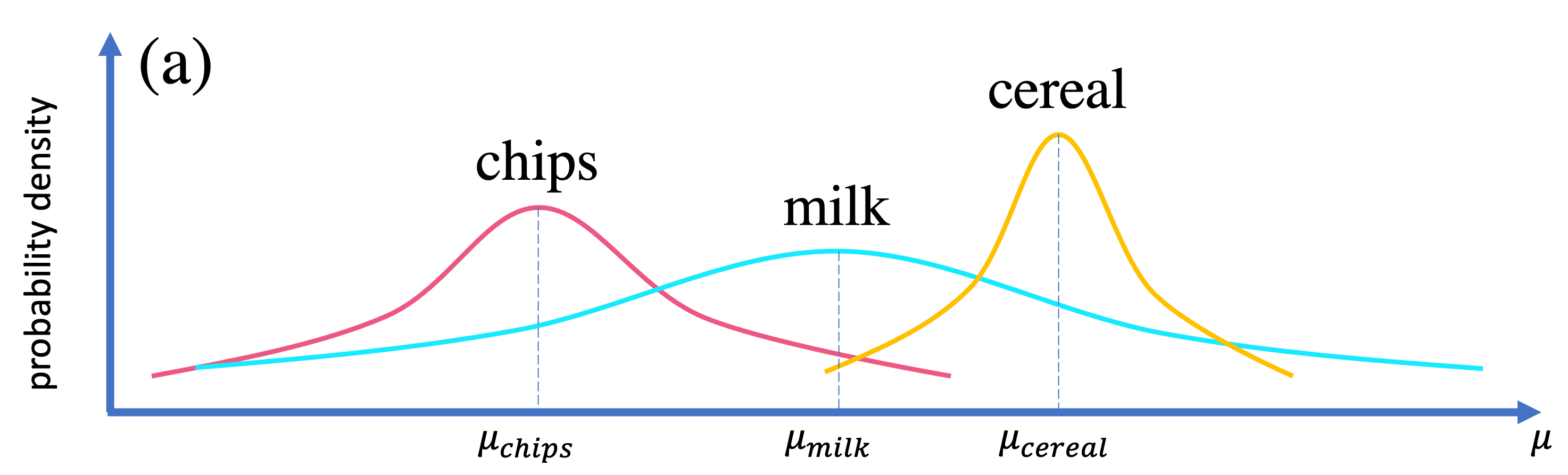}
    \end{minipage}%

    \begin{minipage}[t]{0.48\textwidth}
        \includegraphics[width=\textwidth]{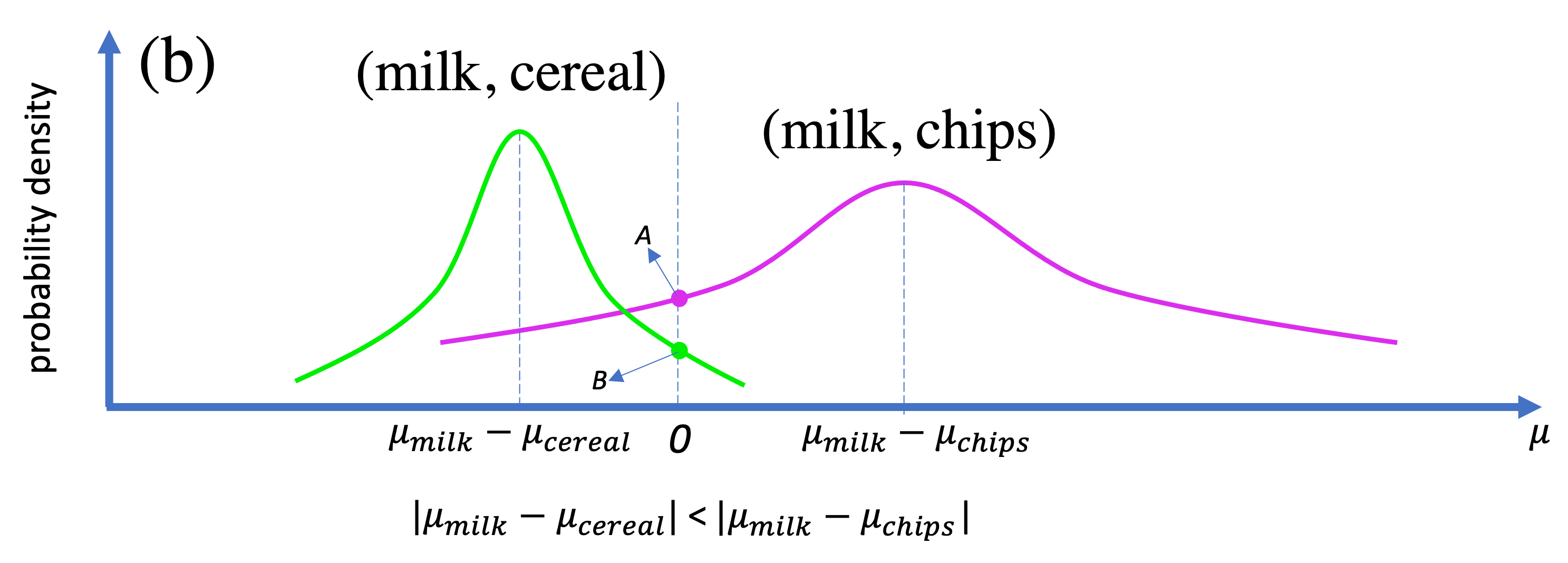}
    \end{minipage}%
    \caption{(a) Examples of Gaussian embeddings (1-D) of items  \texttt{milk}, \texttt{cereal} and \texttt{chips}. (b) Visualization (1-D) of $\mathcal{N}\left(0; \mbox{\boldmath$\mu$}_{milk}-\mbox{\boldmath$\mu$}_{chips},\mbox{\boldmath$\Sigma$}_{milk}+\mbox{\boldmath$\Sigma$}_{chips}\right)$ (the point $A$) and $\mathcal{N}\left(0; \mbox{\boldmath$\mu$}_{milk}-\mbox{\boldmath$\mu$}_{cereal},\mbox{\boldmath$\Sigma$}_{milk}+\mbox{\boldmath$\Sigma$}_{cereal}\right)$ (the point $B$) when \texttt{milk} serves as an query item. While likelihood of observing (\texttt{milk}, \texttt{chips}) could be larger than that of observing (\texttt{milk}, \texttt{cereal}) in the noisy co-purchase records ($A > B$), the correct complementary relationship between \texttt{milk} and \texttt{cereal} is captured by the distance between $\mbox{\boldmath$\mu$}_{milk}$ and $\mbox{\boldmath$\mu$}_{cereal}$.} 
    \label{fig:concept}
\end{figure}

To illustrate the benefit of representing both co-purchase data and items embeddings as Gaussian distributions, let's consider an example from our daily shopping: \texttt{milk}, \texttt{cereal} and \texttt{chips}, in a 1-dimensional embedding space in Figure \ref{fig:concept}.
In Figure \ref{fig:concept}-(a), \texttt{milk} has the largest variance among three items because it is usually a must-buy for many customers and very likely to be co-purchased with other items without complementary relationships. 
\texttt{Cereal} has the smallest variance due to its stable co-purchase behavior with \textit{milk}.
The variance of \texttt{chips} is intermediate because it has some stable combinations such as \texttt{chips dips} while users might also buy them individually as a snack before checkout, which makes it variance relatively larger.
In Figure \ref{fig:concept}-(b), we show the Gaussian distribution of their complementary relationship and highlight the their probability density at zero by the point $A$ for (\texttt{milk}, \texttt{chips}) and the point $B$ for (\texttt{milk}, \texttt{cereal}) when \texttt{milk} serves as the query item.
Because the difference of variances, the Gaussian distribution of the co-purchase for (\texttt{milk}, \texttt{cereal}) shows less variance than that for (\texttt{milk}, \texttt{chips}).
Even though the observed co-purchase records between \texttt{milk} and \texttt{chips} might be more than those between \texttt{milk} and \texttt{cereal}, our model can still capture the correct order of complementary relationships by comparing $|\mathbf{\mu}_{milk} - \mathbf{\mu}_{cereal}|$ and $|\mathbf{\mu}_{milk} - \mathbf{\mu}_{chips}|$. 
However, previous methods using item embeddings might result differently due to the directly fit for co-purchase frequency. 


We follow the paradigm of Skip-gram with Negative Sampling (SGNS) and generate the negative sample $v'$ which is not co-purchased with $q$. Following the margin-based loss \cite{DBLP:journals/corr/VilnisM14}, we construct a max-margin loss function with the margin $\gamma$ in Equation \ref{eq:item-max-margin}: 

\begin{equation}
    \mathcal{L}_{item}(q, v, v') = \max (0, \gamma - \log E(q, v) + \log E(q, v'))
    \label{eq:item-max-margin}
\end{equation}
where 
\begin{align} 
    \log E(q, v) = & -\frac{1}{2} \log \det \left( \mbox{\boldmath$\Sigma$}_{q}+\mbox{\boldmath$\Sigma$}_{v} \right)  - \frac{d}{2} \log (2\pi) \nonumber \\ 
    & -\frac{1}{2}(\mbox{\boldmath$\mu$}_{q}-\mbox{\boldmath$\mu$}_{v})^T\left(\mbox{\boldmath$\Sigma$}_{q}+\mbox{\boldmath$\Sigma$}_{v}\right)^{-1}(\mbox{\boldmath$\mu$}_{q}-\mbox{\boldmath$\mu$}_{v}) \nonumber 
\end{align}


\subsection{Connection to Existing User-Item-level CIRS models} \label{sec:neat_bpr}
Existing works on CIRS such as \cite{DBLP:conf/cikm/WanWLBM18} show that user information will help improve the learning of item-to-item relationship in a collaborative way by introducing the user embedding to the \textbf{Item2Vec} \cite{DBLP:conf/mlsp/BarkanK16} model.
Our model can be extended easily with user information.
Specially, we adapt the advantage of modeling the cohesion of each (\textit{item, item, user}) triplet and modify the BPR loss \cite{DBLP:journals/corr/abs-1205-2618} to model user-item relationship by minimizing the loss function \ref{eq:bpr_u_q} and \ref{eq:bpr_u_r}, where $\sigma(\cdot)$ is the sigmoid function and $q', v'$ represent the negative samples that are not purchased.
These loss functions can be combined together with $\mathcal{L}_{item}(q, v, v')$ to form a new loss function $\mathcal{L}_{item}(q, v, v' | u) = \mathcal{L}_{item}(q, v, v') + \mathcal{L}_{BPR}(u, q, q') + \mathcal{L}_{BPR}(u, v, v')$.

\begin{equation}
        \mathcal{L}_{BPR}(u, q, q') = 1 - \sigma \left( \mbox{\boldmath$\theta$}_u^T \mbox{\boldmath$\mu$}_{q} - \mbox{\boldmath$\theta$}_u^T \mbox{\boldmath$\mu$}_{q'} \right)
        \label{eq:bpr_u_q}
\end{equation}

\begin{equation}
        \mathcal{L}_{BPR}(u, v, v') = 1 - \sigma \left( \mbox{\boldmath$\theta$}_u^T \mbox{\boldmath$\mu$}_{v} - \mbox{\boldmath$\theta$}_u^T \mbox{\boldmath$\mu$}_{v'} \right) 
        \label{eq:bpr_u_r}
\end{equation}


\subsection{Optimization}
Depending on whether we consider $\mathcal{L}_{item}(q, v, v')$ or $\mathcal{L}_{item}(q, v, v' | u)$ for a given co-purchase record $(q, v)$, the final objective function $\mathcal{L}$ can be written in Equation \ref{eq:total_loss}, where $S$ denotes the sampled records for training and $\mathcal{L}_{item}$ could be $\mathcal{L}_{item}(q, v, v')$ or $\mathcal{L}_{item}(q, v, v' | u)$.
We optimize $\mathcal{L}$ by mini-batch Stochastic Gradient Descent.

\begin{equation}
\label{eq:total_loss}
    \mathcal{L} = \sum_{(q, v, v', u) \in S} \mathcal{L}_{item}
\end{equation}

\subsection{Complementary Item Recommendation}
To recommend complementary items, we extract the item Gaussian embeddings and treat the mean vector of each item as its representation under complementary relation.
To mitigate the impact of the vector magnitude when computing the distance between mean vectors for ranking and comparison, we follow \textbf{Item2Vec} \cite{DBLP:conf/mlsp/BarkanK16} and \textbf{Triple2Vec} \cite{DBLP:conf/cikm/WanWLBM18} and use the cosine similarity between two items' mean vectors to represent the relevance of the complementary relationship.

%% file: labels.tex
\section{Trustworthy Evaluation} \label{label_generation}
Although we have addressed the label noise issue in the modeling step by considering the co-purchase data as a Gaussian distribution with item Gaussian embeddings, label noise will impact the evaluation accuracy as well for result reporting purpose. In this section, a trustworthy evaluation is developed to exam the models with high quality labels generated from an independence test-based method. Note that this evaluation does not require extra information (item description, co-view data, etc.) for creating the high quality labels. 



Inspired from the definition of complementary items, we treat the purchase of an individual item $v$ as a random variable from a Bernoulli distribution $Y_{v} \sim \textbf{Bernoulli}(p_{v})$, and study the independence between two items' purchase to surface the item pairs which are co-purchased dependently. 
Pearson's chi-squared test is suitable for this task, as it can assess whether observations consisting of measures on two variables, expressed in a contingency table, are independent of each other \cite{pearson1900x}. 
Given two co-purchased items $v_i$ and $v_j$, we define the 2-by-2  contingency table (Table \ref{tab:contingency_table}) for the observations of the purchase event between $v_i$ and $v_j$  with the 1 degree of freedom.
Let $N$ denote the total number of observed co-purchase records 
in the evaluation dataset.
$\mathcal{F}_{v_i}$ ($\mathcal{F}_{v_j}$) represents the frequency of co-purchases including the item $v_i$ ($v_j$)
and $O_i$ represents the observed frequency of different purchase events defined in Table \ref{tab:contingency_table}. 
Typically, $O_1$ represents the observed co-purchases of ($v_i$, $v_j$). 
Following the definition of $\mathcal{F}_{v_i}$ and $\mathcal{F}_{v_j}$, we can compute that $O_2 = \mathcal{F}_{v_j} - O_1$, $O_3 = \mathcal{F}_{v_i} - O_1$ and $O_4 = N - O_1 - O_2 - O_3 = N - \mathcal{F}_{v_i} - \mathcal{F}_{v_j} + O_1$.

Without any knowledge of item complementary relationships, we assume that each pair of co-purchased items, $v_i$ and $v_j$, are independent (the null hypothesis in our test $H_0$).
The alternative hypothesis $H_a$ is that they are purchased dependently.
We can compute the estimated frequency for each purchase event $E_i$ based on the independence assumption by Table \ref{tab:contingency_table_expectation}.
Following the Chi-squared test, we can compute the value of the Chi-squared statistics $\mathcal{X}^2 = \sum_{i=1}^4 \frac{(O_i - E_i)^2}{E_i}$ which is used to determine the significance (p-value) by comparing to a Chi-squared distribution with one degree of freedom.
Item pairs which pass the Chi-squared test mean that their co-purchase are dependent.

Further more, we need to determine if the dependency of a co-purchased item pair is positive or negative. 
To achieve this, we require that the observed co-purchase frequency of an item pair should be larger than the expected frequency under independence assumption, $O_1 > E_1$, if a co-purchased item pair has a positive dependency.
With a predefined p-value for the statistic significance, we can create the high quality co-purchase labels for evaluations.
For clarity, we denote the item pairs which pass the Chi-squared test and $O_1 > E_1$ as the \textbf{positively-dependent item pairs} and the item pairs which pass the Chi-squared test and $O_1 <= E_1$ as the \textbf{negatively-dependent item pairs} in the rest of our paper.
We summarize the algorithm of generating the trustworthy labels for evaluation in Algorithm \ref{algo:label_generation}.

\begin{algorithm}
    \caption{Trustworthy Label Generation for Evaluation}
    \label{algo:label_generation}
    
    \begin{algorithmic}[1]
    \REQUIRE a transaction set $\mathcal{B}$, an empty hashtable $\Psi$, $\mathcal{X}^2$ threshold $t_{\mathcal{X}^2}$ for a p-value;
    
    \ENSURE ~~\\    
    \FOR{each transaction $b$ in $\mathcal{B}$}
        \STATE sample co-purchase item pairs $(v_i, v_j)$ from each transaction $b \in \mathcal{B}$, $v_i \neq v_j$;
        \STATE compute the frequency of purchasing $(v_i, v_j)$ together and store the frequency in $\Psi$, i.e., $\Psi[(v_i, v_j)]$ represents the co-purchase frequency of $(v_i, v_j)$;
    \ENDFOR
    
    \STATE set $N = \sum_{(v_i, v_j)} \Psi[(v_i, v_j)]$;
    \STATE set $\mathcal{F}_{v_i} = \sum_{(v_k, v_j), v_k = v_i} \Psi[(v_k, v_j)]$;
    \STATE set $\mathcal{F}_{v_j} = \sum_{(v_i, v_k), v_k = v_j} \Psi[(v_i, v_k)]$;
    
    \FOR{each $(v_i, v_j)$ stored in $\Psi$}
        \STATE compute the 2-by-2 contingency table by $\Psi[(v_i, v_j)]$, $N, \mathcal{F}_{v_i}$ and $\mathcal{F}_{v_j}$ based on Table \ref{tab:contingency_table};
        \STATE compute the table of expected value based on Table \ref{tab:contingency_table_expectation};
        \STATE compute $\mathcal{X}^2_{(v_i, v_j)} = \sum_{i=1}^4 \frac{(O_i - E_i)^2}{E_i}$
        \IF{$\mathcal{X}^2_{(v_i, v_j)} > t_{\mathcal{X}^2}$ and $O_1 > E_1$}
        \STATE mark $(v_i, v_j)$ as a qualified co-purchase label for evaluation
        \ENDIF
    \ENDFOR

    \end{algorithmic}
\end{algorithm}


\begin{table}
          \footnotesize
          \caption{2-by-2  contingency table of 4 different purchase events between $v_i$ and $v_j$.}
          \label{tab:contingency_table}
          \centering 
          \begin{tabular}{c|c|c|c}
            \hline
             &  $Y_{v_i} = 1$ & $Y_{v_i} = 0$ & \textbf{SUM}\\
            \hline
            $Y_{v_j} = 1$ & \makecell[c]{ $O_{1}$ = frequency of \\ observed  co-purchase \\ ($v_i, v_j$)} & \makecell[c]{$O_{2}$ = frequency of \\ observed co-purchase \\ of $v_j$ with all items\textbackslash $v_i$} & $\mathcal{F}_{v_j}$  \\
            \hline
            $Y_{v_j} = 0$ & \makecell[c]{$O_{3}$ = frequency of  \\ observed co-purchase \\ of $v_i$ with all items\textbackslash $v_j$} &  \makecell[c]{$O_{4}$  = frequency of \\ observed  co-purchase \\  w/o ($v_i$, $v_j$) } & $N$ - $\mathcal{F}_{v_j}$ \\
            \hline 
            \textbf{SUM} & $\mathcal{F}_{v_i}$ & $N$ - $\mathcal{F}_{v_i}$ & $N$ \\
            \hline 
        \end{tabular}
\end{table}

\begin{table}
          \footnotesize
          \caption{expected value of 4 different purchase events between $v_i$ and $v_j$.}
          \label{tab:contingency_table_expectation}
          \centering 
          \begin{tabular}{c|c|c}
            \hline
             &  $Y_{v_i} = 1$ & $Y_{v_i} = 0$  \\
            \hline
            $Y_{v_j} = 1$ & \makecell[c]{ $E_{1} = \frac{\mathcal{F}_{v_i} \cdot \mathcal{F}_{v_j}}{N}$} 
            & \makecell[c]{$E_{2} = \frac{(N-\mathcal{F}_{v_i}) \cdot \mathcal{F}_{v_j}}{N}$} \\
            \hline
            $Y_{v_j} = 0$ & \makecell[c]{$E_{3} = \frac{\mathcal{F}_{v_i} \cdot (N - \mathcal{F}_{v_j})}{N}$} &  \makecell[c]{$E_{4} = \frac{(N-\mathcal{F}_{v_i})\cdot(N-\mathcal{F}_{v_j})}{N} $ } \\
            \hline 
        \end{tabular}
\end{table}

%% file: experiments.tex
\section{EXPERIMENTS} \label{sec:experiments}
In this section, we study \NEAT by comparing it with the state-of-the-art baselines on the real-world datasets.

\subsection{Dataset}
For the publicly available dataset of raw transactions, we consider \textit{Instacart} dataset (INS) published by \cite{instacart}. 
The date of each order in this dateset is not provided but the sequence of transactions by each user is available.
Items in each transaction are sorted by their purchase orders and the item-types are also provided in \textit{Instacart} by the aisles. 
\textit{Instacart} dataset has 134 aisles from 21 departments and 3.3 million transactions, which is small compared with the real-world applications with more item-types and larger volume of transactions.
We use the default train (INS-T) and test (INS-E) split provided by \textit{Instacart} dataset. 
To further study the model performance, we collect a proprietary dataset (WMT) with a larger scale from \textit{Walmart} e-Commerce platform (www.walmart.com) following the same format of \textit{Instacart}, where the sequence order of transactions are kept and the order of purchases in the same sequence is also preserved.
For WMT dataset, we randomly sample $15.2$ million transactions from the past 6-month history data and keep the latest $1.2$ million transactions as our test dataset (WMT-E). 
The rest of $14$ million transactions are used for training (WMT-T).
Similar to INS dataset, we collect the item categories based on the taxonomy of \textit{Walmart} platform.
Co-purchase records are created from INS and WMT dataset respectively to serve model training and label generation for evaluation.
Table \ref{tab:dataset} summarizes the statistics of the INS and WMT datasets.

\subsection{Label Generation for Training and Evaluation} 
We follow the steps in \ref{sample_co-purchase} to collect all the co-purchase records for training from the training set. 
To improve the quality of labels for model training, we remove labels selected in the previous steps where two items are from the same aisles (for INS dataset) or the same category (for WMT dataset) to remove similar items.
This is similar to the strategy 
used in \cite{DBLP:conf/cikm/HaoZLDFSW20} when co-view data are not available because items from the same aisle or category are similar and are likely to be co-viewed for substitution. 

For evaluation, we create the trustworthy labels following Section \ref{label_generation}
under different p-value = $\{0.05, 0.01, 0.001\}$.
We conduct the experiments on these unique labels for evaluation. 
As we mentioned previously, even these labels are with high quality, it is not practical to be used for training purpose, due to the limited coverage in item space. Table \ref{tab:labels} summarizes the number of unique labels of the INS and WMT datasets.

\begin{table}
          \footnotesize
          \caption{Data Description of WMT and INS dataset}
          \label{tab:dataset}
          \centering 
          \begin{tabular}{lcccc}
            \hline
             &  INS-T & INS-E & WMT-T & WMT-E  \\
            \hline 
            \# Transactions & 3,214,874 & 131,209 & $\sim$14 m & $\sim$1.2 m  \\
            \# Items  & 49,677 & -- & $\sim$0.1 m & -- \\
            \# Categories/Aisles  & 134 & -- & $\sim$850 & -- \\
            \# Users  &  206,209 & 131,209 & $\sim$0.7 m & $\sim$0.4 m\\
            \hline
        \end{tabular}
\end{table}

\begin{table}
          \footnotesize
          \caption{Number of Labels in INS-E and WMT-E}
          \label{tab:labels}
          \centering 
          \begin{tabular}{lccc}
            \hline
             &  p-value = 0.05 & p-value = 0.01 & p-value = 0.001  \\
            \hline 
            INS-E & 7,961 & 6,077 & 4,752 \\ 
            WMT-E & 78,719 & 53,119 & 36,251 \\
            \hline
        \end{tabular}
\end{table}

\subsection{Experiment Setup}
\subsubsection{\textbf{Baselines}}
To evaluate the effectiveness of applying Gaussian distribution on co-purchase data and item embeddings, we compare \NEAT with the following state-of-the-art baselines:
\begin{itemize}
    \item  \textbf{Collaborative Filtering (\textbf{CF})} \cite{DBLP:conf/icdm/HuKV08}: an item recommendation model which factorizes the user-item. 
    \item \textbf{Bayesian Personalized Ranking (\textbf{BPRMF})} \cite{DBLP:journals/corr/abs-1205-2618}: an item recommendation model which factorizes the user-item implicit feedback from raw transactions by approximately optimizing the AUC ranking metric. 
    \item \textbf{Item2Vec} \cite{DBLP:conf/mlsp/BarkanK16}: the first model that learns vector representations of items via SGNS and optimizes the similarity between item vectors for co-purchase data. 
    It can be used to model item complementarity by considering co-purchase records of item pairs as input. 
    As aforementioned, most of the CIRS models can be viewed as \textbf{Item2Vec} plus auxiliary information such as graph, context and multi-modal data source.
    In our work, we focus on modeling item complementary relationship rather than the advantage of incorporating such auxiliary information into models.
    Hence, we choose this model as the baseline to represent other item-to-item CIRS models for a fair comparison.
    \item \textbf{Triple2Vec} \cite{DBLP:conf/cikm/WanWLBM18}: this state-of-the-art model learns vector representations of item and user, and considers the triplet interaction between a user and her/his co-purchased item pair for complementarity.
    It can be viewed as an extension of \textbf{Item2Vec} with user embeddings. 
\end{itemize}
Besides, we also consider two popularity-based baselines: \textbf{Popular Item (Pop)} and \textbf{Popular Co-purchase (\textbf{PopCo})}. 
In \textbf{Pop}, the complementary item recommendations for the query item are the most popular items globally. 
In \textbf{PopCo}, we take the query item's popular co-purchased items as the complementary item recommendations.

\subsubsection{\NEAT Variants} \label{sec:facirs_variant}
Depending on whether incorporating the user-item level collaborative learning into the model, we develop two variants of our model:
\begin{itemize}
    \item \NEAT:
    This model is trained by optimizing $\mathcal{L}$ with $\mathcal{L}_{item} = \mathcal{L}_{item}(q, v, v')$ to model the item-level complementary co-purchase signals.
     \item \NEATe+bpr:
    In addition to the item-level complementary signals, this model is trained by optimizing $\mathcal{L}$ with $\mathcal{L}_{item} = \mathcal{L}_{item}(q, v, v'|u)$ (see section \ref{sec:neat_bpr}) to further model the user-item level collaborative learning for complementary signals.
\end{itemize}

\subsubsection{\textbf{Implementation Details}}
For simplicity, we set the covariance matrix in the \NEAT model to be spherical.
The margin $\gamma$ in Equation \ref{eq:item-max-margin} is set to be $0.5$ for the computation of Hit-Rate (HR) and Normalized Discounted Cumulative Gain (NDCG).
We applied the following settings for all models in the experiments, unless it is specified: 
the dimension of the item embeddings are set to be $100$, 
the window size for sampling co-purchased items is set to be $5$, 
and all models are trained for $5$ epochs.  
For \textbf{Item2Vec}, \textbf{Triple2Vec} and our model, the batch size is $128$, with the initial learning rate of $0.05$ and the mini-batch Stochastic Gradient Descent (SGD) optimizer. 
We follow the skip-gram training paradigm and set the number of negative sample to $5$ during training.

\subsection{Study on Labels for Evaluation}
\subsubsection{Label Quality}
In this section, we present the study of the trustworthy label generation method and show its effectiveness.
There are three major concerns of data labeling: coverage, consistency and accuracy. 

\noindent \textbf{Coverage}: a good data labeling method should have enough coverage on the representative patterns of the dataset. 
In our case, the label generation should show a good coverage of different item categories and departments instead of being biased to few item categories. 
To illustrate the coverage of our label generation method, we focus on the department level without the loss of generality and readability and compute the distribution of labels over different departments for the INS-E dataset in Table \ref{tab:labe_dist_over_dept}. 
Compared with the distribution of total co-purchase records from the INS-E dataset, our labels show similar distributions over all departments.
We notice that the Pets department is not covered by our labels. 
This is because most of the raw co-purchase records with pet-related items also consist of non-pet-related items like grocery in the INS-E dataset, which are not complementary.
The label distribution over departments indicates that our method is not biased to a certain department and covers complementary signals of item purchase behaviors under various departments.

\begin{table}[]
\centering
\footnotesize
\caption{Distribution of Labels over Departments of INS-E Dataset}
\label{tab:labe_dist_over_dept}
\begin{tabular}{lcccc}
\hline
\small{\textbf{Department}} & \small{Total} & \small{$p=0.05$} &  $p=0.01$ & $p=0.001$ 
\\
\hline
alcohol  & 0.370\% & 0.339\% & 0.411\% & 0.505\% \\
babies & 1.451\% & 0.364\% & 0.378\% & 0.337\% \\
bakery & 4.462\% & 4.459\% & 4.048\% & 3.788\% \\
beverages & 9.437\% & 8.504\% & 8.672\% & 9.007\% \\
breakfast & 2.796\% & 1.306\% & 1.349\% & 1.410\% \\
bulk & 0.100\% & 0.100\% & 0.115\% & 0.126\% \\
canned goods & 4.001\% & 2.927\% & 2.781\% & 2.399\% \\
dairy eggs & 17.101\% & 23.678\% & 22.100\% & 20.623\% \\
deli & 3.594\% & 2.487\% & 2.403\% & 2.273\% \\
dry goods pasta & 3.559\% & 1.859\% & 1.678\% & 1.599\% \\
frozen & 8.715\% & 3.944\% & 3.966\% & 4.167\% \\
household & 3.016\% & 0.867\% & 0.889\% & 0.989\% \\
international & 1.138\% & 0.603\% & 0.675\% & 0.526\% \\
meat seafood & 2.474\% & 2.613\% & 2.271\% & 2.041\% \\
missing & 0.759\% & 1.017\% & 1.152\% & 1.305\% \\
other & 0.168\% & 0.038\% & 0.049\% & 0.042\% \\
pantry & 6.720\% & 2.927\% & 2.946\% & 2.736\% \\
personal care & 1.772\% & 0.113\% & 0.115\% & 0.147\% \\
pets & 0.484\% & 0.000\% & 0.000\% & 0.000\% \\
produce & 17.369\% & 30.511\% & 31.019\% & 31.587\% \\
snacks & 10.512\% & 11.343\% & 12.983\% & 14.394\% \\
\hline
\textbf{SUM} & 6249077 & 7961 & 6077 & 4752 \\ 
\hline
\end{tabular}
\end{table}

\begin{figure*}[t]
    \begin{minipage}[t]{0.327\textwidth}
    \centering
    \includegraphics[width=\linewidth]{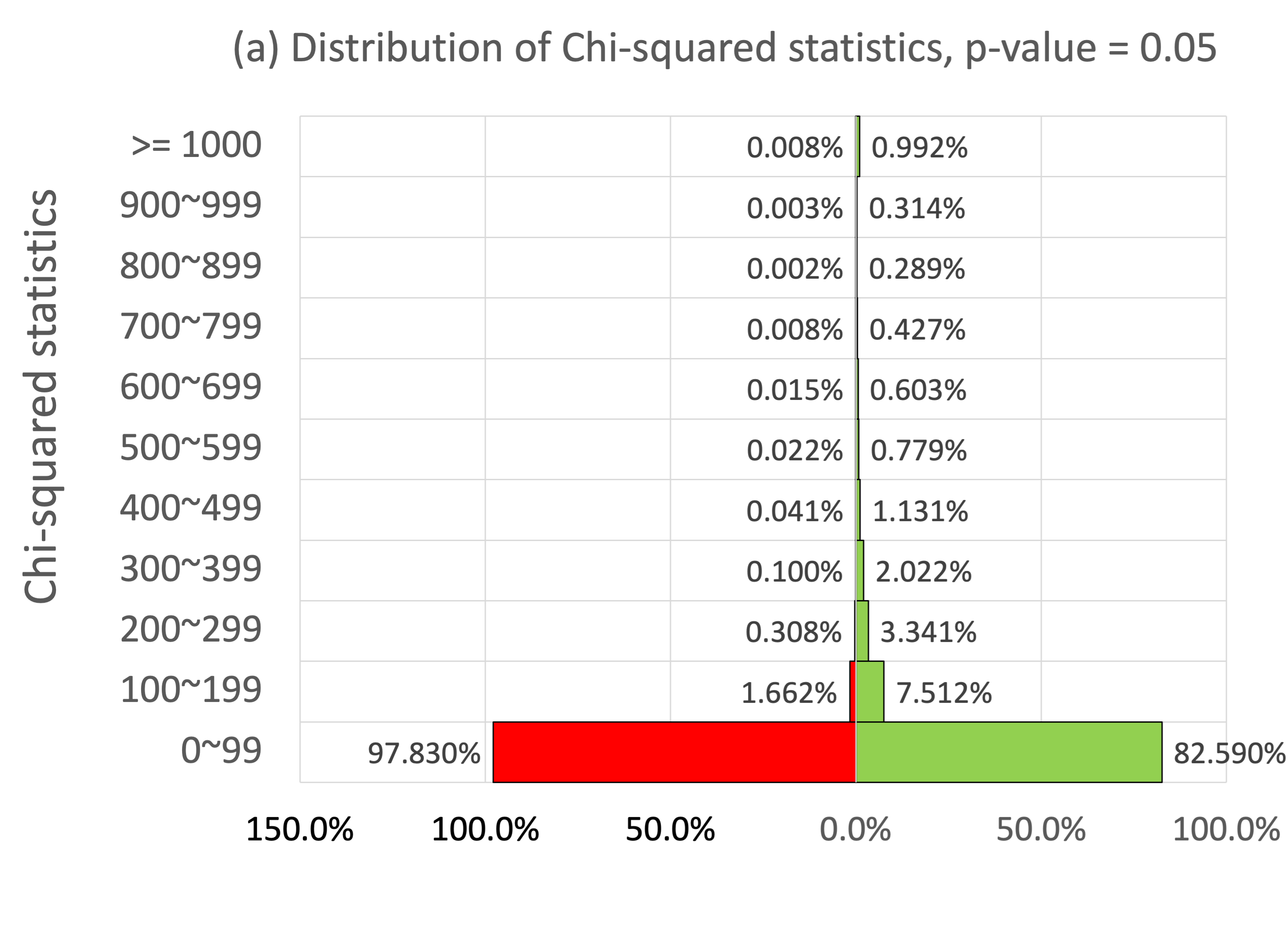}
    \end{minipage}%
    \hfill
    \begin{minipage}[t]{0.33\linewidth}
    \centering
      \includegraphics[width=\linewidth]{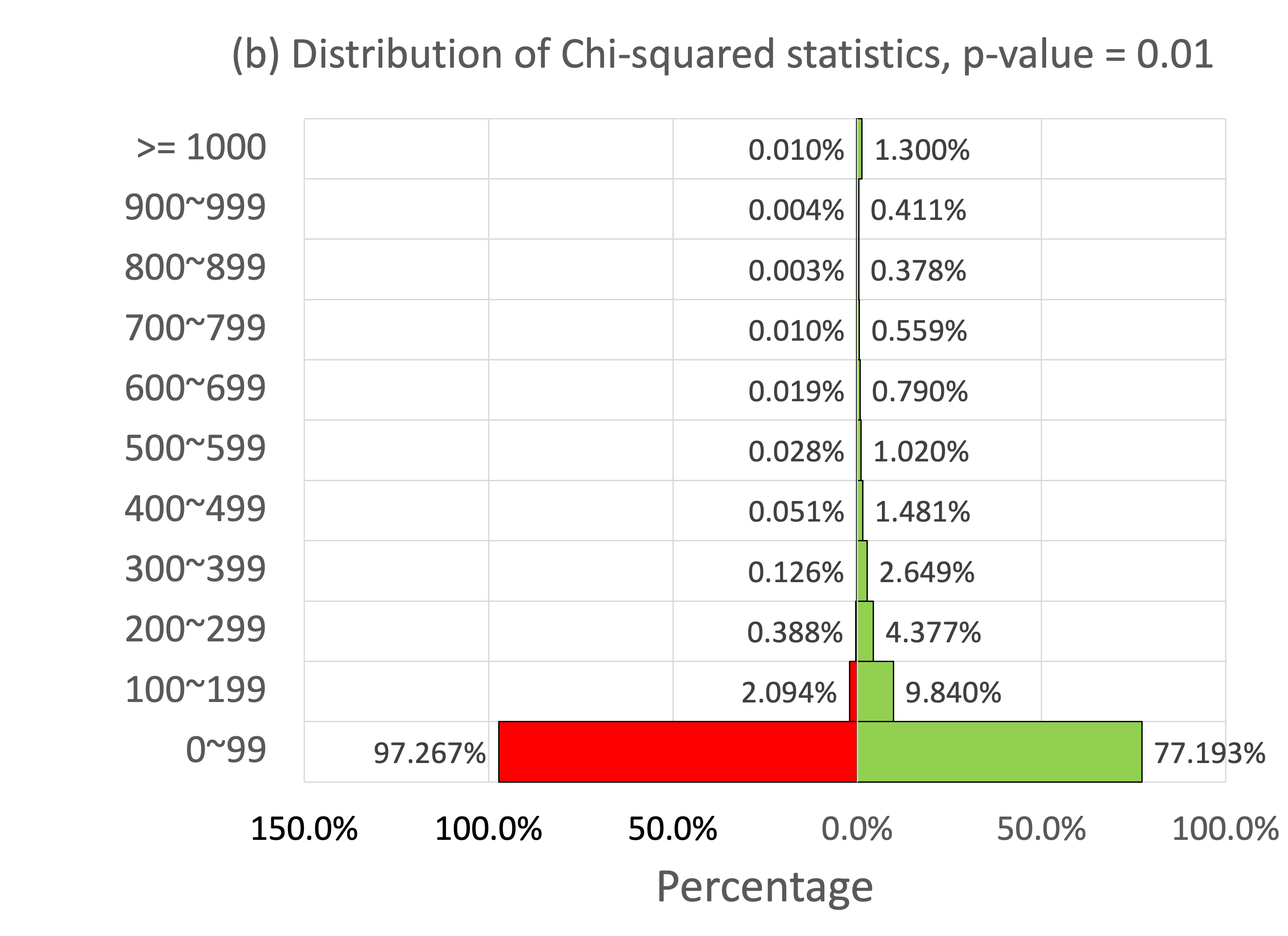}
    \end{minipage}    
    \hfill 
    \hfill
    \begin{minipage}[t]{0.33\linewidth}
    \centering
      \includegraphics[width=\linewidth]{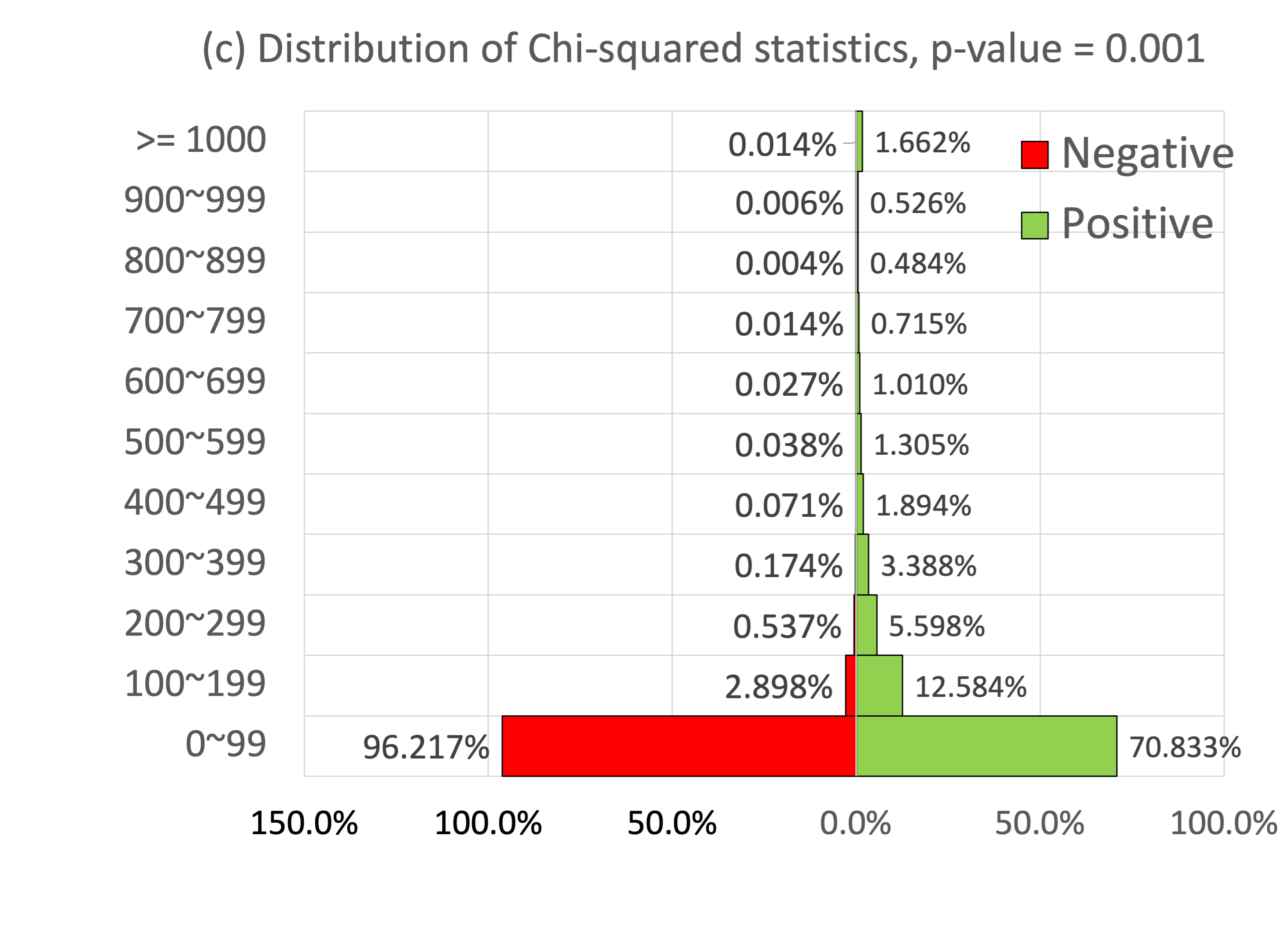}
    \end{minipage}    
    \caption{Distribution of $\mathcal{X}^2$ of both positively dependent labels (in green) and negatively dependent labels (in red) with p-value = $\{0.05, 0.01, 0.001\}$.}
    \label{fig:chi_squared_dist}
\end{figure*}

\noindent \textbf{Consistency}: by the design of our label generation method, the percentage of complementary labels should increase as the p-value decreases. 
To show such a consistency, we plot the distribution of $\mathcal{X}^2$ statistics for each item pair which passes the test for a given p-value for both positively dependent item pairs ($O_1 > E_1$) and negatively dependent item pairs ($O_1 <= E_1$) in Figure \ref{fig:chi_squared_dist}.
We see that while most of the labels (both negative and positive) are with $\mathcal{X}^2$ statistics between $0$ and $99$, the percentage of positively dependent item pairs with higher $\mathcal{X}^2$ statistics has a larger lift as the p-value decreases compared with the negatively dependent item pairs.
Because we use the positively dependent item pairs as the labels for evaluation, this consistency between the increase of more complementary labels and the decrease of p-value indicates that raising the significance level by p-value can further concentrate the co-purchase labels with positively dependence (complementary relationships).

\noindent \textbf{Accuracy}: we provide a case study of the positively dependent item pairs (our labels) and the negatively dependent item pairs to show that our model can provide more accurate labels for evaluation in Table \ref{tab:ins_positive} and \ref{tab:ins_negative}.
Note that the chi-squared statistics $\mathcal{X}^2$ should be not smaller than $10.83$ for p-value = 0.001.
Both positive and negative item pairs show large enough chi-squared statistics.
While the positive labels are showing clear complementary relationships, e.g., syrup for waffle, hot dog buns for hot dog, and kitchen bag for laundry-related items for household, the negative labels reflect the noise in the co-purchase records even though they pass the Chi-squared test.
Most of the co-purchased items in the negative labels are fruits like Banana, which are the popular items in the INS dataset. 
See examples of top-20 popular items in the INS dataset in Table \ref{tab:pop_top_20}.
The comparison between the positive labels and the negative labels indicates that our label generation method can surface more complementary labels while suppressing the noise in the co-purchase records.
\begin{table}[]
    \centering
    \caption{Positively-dependent Item pair, INS dataset with p-value = 0.001}
    \label{tab:ins_positive}
    \begin{tabular}{ccc}
    \hline
        \small\textbf{Query Item} & \small\textbf{Co-purchased Item}  & \small\textbf{$\mathcal{X}^2$}  \\
        \hline
        Beef Hot Dogs & Classic Hot Dog Buns & 5084.536\\ 
        Everything Bagels & Whipped Cream Cheese & 85.501 \\ 
        Thin \& Light Tortilla Chips & Medium Salsa Roja & 239.958 \\ 
        Eggo Homestyle Waffles & Original Syrup & 170.825\\ 
        Cherrios Honey Nut (cereal) & Reduced Fat 2\% Milk & 62.804\\ 
        Green Curry Paste & Organic Coconut Milk & 51.005\\ 
        Plain Mini Bagels & \makecell[c]{Philadelphia Cream \\ Cheese Spread} & 33.513\\ \
        Stand 'n Stuff Taco Shells & Original Taco Seasoning Mix & 20.774\\ 
        Snack Bags (food storage) & Sandwich Bags (food storage) & 106.078 \\ 
        Fabric Softener Dryer Sheet & \makecell[c]{Tall Kitchen Bag \\ With Febreze Odor Shield} & 1414.015\\ 
      \hline
    \end{tabular}
\end{table}

\begin{table}[]
    \centering
    \caption{Negatively-dependent Item pair, INS dataset with p-value = 0.001}
    \label{tab:ins_negative}
    \begin{tabular}{ccc}
    \hline
        \small\textbf{Query Item} & \small\textbf{Co-purchased Item}  & \small\textbf{$\mathcal{X}^2$} \\
        \hline
        \makecell[c]{Organic Sea Salt \\ Roasted Seaweed Snacks} & Banana & 108.817 \\ 
        Free \& Clear Unscented Baby Wipes & Large Lemon & 61.033\\ 
        \makecell[c]{Naturals Savory Turkey \\Breakfast Sausage} & Strawberries & 18.558 \\ 
        Gluten Free Whole Grain Bread & Large Lemon & 52.104 \\ 
        Eggo Homestyle Waffles & Organic Cucumber & 42.681 \\ 
        Naturals Chicken Nuggets & Organic Avocado & 60.222 \\ 
        Cheerios Honey Nut (cereal) & Jalapeno Peppers & 33.853 \\ 
        Everything Bagels & Organic Strawberries & 35.512 \\ 
        Taco Seasoning & Organic Raspberries & 53.735\\ 
        Laundry Detergent Free \& Clear & Banana & 16.293 \\ 
      \hline
    \end{tabular}
\end{table}

\begin{table}[]
    \centering
    \caption{Top-20 Globally Popular Items, INS dataset}
    \label{tab:pop_top_20}
    \begin{tabular}{cc}
    \hline
        \small \textbf{Rank} & \small \textbf{Item} \\ 
        \hline 
        1 & Banana \\ 
        2 & Bag of Organic Bananas \\ 
        3 & Organic Strawberries \\ 
        4 & Organic Baby Spinach \\ 
        5 & Organic Hass Avocado \\ 
        6 & Organic Avocado \\ 
        7 & Large Lemon \\ 
        8 & Strawberries \\ 
        9 & Limes \\ 
        10 & Organic Whole Milk \\ 
        11 & Organic Raspberries \\ 
        12 & Organic Yellow Onion \\ 
        13 & Organic Garlic \\ 
        14 & Organic Zucchini \\ 
        15 & Organic Blueberries \\ 
        16 & Cucumber Kirby \\ 
        17 & Organic Fuji Apple \\ 
        18 & Organic Lemon \\ 
        19 & Apple Honeycrisp Organic \\ 
        20 & Organic Grape Tomatoes \\ 
        \hline
    \end{tabular}
    
\end{table}

\subsection{Evaluation on Item-level Co-purchase Data}
\noindent\textbf{Evaluation metrics}: 
We mainly focus on HitRate (HR@$K$) and NDCG@$K$ of evaluation.
Given the query item $q$, we consider the top-$K$ recommendations $R_q$ has a hit on the test co-purchase record $(q, v)$ if $v \in R_q$: $ \text{HR@}K = 
            \begin{cases}
                1 ,& \text{if } v \in R_q\\
                0,              & \text{otherwise}
            \end{cases} $. 
For NDCG@$K$, we consider the binary relevance score and define it as $ \text{NDCG@}K = 
             \begin{cases}
                \frac{1}{\log_2 (1 + rank_v)} ,& \text{if } v \in R_q\\
                0,              & \text{otherwise}
            \end{cases} $.

To evaluate the ability to surface complementary recommendations from the noisy co-purchase data, we firstly generate the recall set by taking the top-$K$ most co-purchased items for the query item in the training data, rather than a sampled item set in which each ground truth item in the test set is paired with a few (e.g., 100) randomly sampled negative items \cite{DBLP:conf/www/HeLZNHC17}\cite{DBLP:conf/icdm/KangM18}\cite{DBLP:conf/cikm/SunLWPLOJ19}\cite{DBLP:conf/wsdm/TangW18}.
We report the average score over the co-purchase records for HR@$K$ and NDCG@$K$, $K = \{1, 3, 5, 10, 20\}$.


\begin{table*}[]
    \centering
    \caption{INS Labels, p-value = 0.05}
    \label{tab:metric_1}
    \begin{tabular}{l|cc|cc|cc|cc|cc}
    \hline
    & \textbf{HR@1} & \textbf{NDCG@1} & \textbf{HR@3} & \textbf{NDCG@3} & \textbf{HR@5} & \textbf{NDCG@5} & \textbf{HR@10} & \textbf{NDCG@10} & \textbf{HR@20} & \textbf{NDCG@20} \\ 
    \hline
\textbf{Pop}  & 0.0000 & 0.0000 & 0.0000 & 0.0000 & 0.0001 & 0.0001 & 0.0014 & 0.0005 & 0.0035 & 0.0010 \\
\textbf{PopCo} & 0.0122 & 0.0122 & 0.0437 & 0.0303 & 0.0734 & 0.0425 & 0.1334 & 0.0617 & 0.2168 & 0.0826 \\
\hdashline
\textbf{CF} & 0.0087 & 0.0087 & 0.0245 & 0.0176 & 0.0396 & 0.0238 & 0.0765 & 0.0355 & 0.1516 & 0.0543 \\
\textbf{BPRMF} & 0.0067 & 0.0067 & 0.0225 & 0.0155 & 0.0368 & 0.0214 & 0.0720 & 0.0326 & 0.1467 & 0.0512 \\
\textbf{Item2Vec} & 0.0196 & 0.0196 & 0.0484 & 0.0360 & 0.0746 & 0.0468 & 0.1271 & 0.0636 & 0.2231 & 0.0876 \\
\textbf{Triple2Vec} & 0.0221 & 0.0221 & 0.0541 & 0.0403 & 0.0813 & 0.0514 & 0.1325 & 0.0678 & 0.2110 & 0.0874 \\
\textbf{NEAT} & \textbf{0.0252} & \textbf{0.0252} & \textbf{0.0633} & \textbf{0.0468} & \textbf{0.0970} & \textbf{0.0606} & 0.1574 & 0.0798 & \textbf{0.2593} & \textbf{0.1054} \\
\textbf{NEAT}+bpr & 0.0249 & 0.0249 & 0.0628 & 0.0464 & 0.0927 & 0.0586 & \textbf{0.1628} & \textbf{0.0811} & 0.2591 & 0.1053 \\
\hline
    \end{tabular}
\end{table*}

\begin{table*}[]
    \centering
    \caption{INS Labels,  p-value = 0.01}
    \label{tab:metric_2}
    \begin{tabular}{l|cc|cc|cc|cc|cc}
    \hline
& \textbf{HR@1} & \textbf{NDCG@1} & \textbf{HR@3} & \textbf{NDCG@3} & \textbf{HR@5} & \textbf{NDCG@5} & \textbf{HR@10} & \textbf{NDCG@10} & \textbf{HR@20} & \textbf{NDCG@20} \\ 
\hline
\textbf{Pop}  & 0.0000 & 0.0000 & 0.0000 & 0.0000 & 0.0000 & 0.0000 & 0.0015 & 0.0005 & 0.0023 & 0.0007 \\ 
\textbf{PopCo} & 0.0155 & 0.0155 & 0.0541 & 0.0378 & 0.0900 & 0.0526 & 0.1593 & 0.0747 & 0.2587 & 0.0996 \\ 
\hdashline
\textbf{CF} & 0.0100 & 0.0100 & 0.0276 & 0.0200 & 0.0443 & 0.0268 & 0.0849 & 0.0397 & 0.1711 & 0.0612 \\ 
\textbf{BPRMF} & 0.0076 & 0.0076 & 0.0262 & 0.0180 & 0.0415 & 0.0243 & 0.0819 & 0.0372 & 0.1654 & 0.0580 \\ 
\textbf{Item2Vec} & 0.0230 & 0.0230 & 0.0559 & 0.0418 & 0.0859 & 0.0541 & 0.1450 & 0.0729 & 0.2549 & 0.1004 \\ 
\textbf{Triple2Vec} & 0.0253 & 0.0253 & 0.0635 & 0.0472 & 0.0931 & 0.0593 & 0.1502 & 0.0775 & 0.2391 & 0.0998 \\ 
\textbf{NEAT} & \textbf{0.0293} & \textbf{0.0293} & \textbf{0.0734} & \textbf{0.0543} & \textbf{0.1121} & \textbf{0.0701} & 0.1833 & 0.0928 & 0.2998 & 0.1221 \\ 
\textbf{NEAT}+bpr & 0.0286 & 0.0286 & 0.0732 & 0.0540 & 0.1084 & 0.0684 & \textbf{0.1899} & \textbf{0.0945} & \textbf{0.3011} & \textbf{0.1224} \\ 
\hline
    \end{tabular}
\end{table*}

\begin{table*}[]
    \centering
    \caption{INS Labels, p-value = 0.001}
    \label{tab:metric_3}
    \begin{tabular}{l|cc|cc|cc|cc|cc}
    \hline
& \textbf{HR@1} & \textbf{NDCG@1} & \textbf{HR@3} & \textbf{NDCG@3} & \textbf{HR@5} & \textbf{NDCG@5} & \textbf{HR@10} & \textbf{NDCG@10} & \textbf{HR@20} & \textbf{NDCG@20} \\ 
\hline
\textbf{Pop}  & 0.0000 & 0.0000 & 0.0000 & 0.0000 & 0.0000 & 0.0000 & 0.0013 & 0.0004 & 0.0023 & 0.0007 \\ 
\textbf{PopCo} & 0.0189 & 0.0189 & 0.0640 & 0.0450 & 0.1048 & 0.0618 & 0.1833 & 0.0869 & 0.2963 & 0.1152 \\ 
\hdashline
\textbf{CF} & 0.0112 & 0.0112 & 0.0316 & 0.0227 & 0.0499 & 0.0302 & 0.0943 & 0.0443 & 0.1896 & 0.0681 \\ 
\textbf{BPRMF} & 0.0082 & 0.0082 & 0.0286 & 0.0197 & 0.0452 & 0.0266 & 0.0922 & 0.0415 & 0.1841 & 0.0645 \\ 
\textbf{Item2Vec} & 0.0265 & 0.0265 & 0.0623 & 0.0468 & 0.0962 & 0.0607 & 0.1616 & 0.0816 & 0.2870 & 0.1129 \\ 
\textbf{Triple2Vec} & 0.0276 & 0.0276 & 0.0711 & 0.0525 & 0.1040 & 0.0659 & 0.1681 & 0.0864 & 0.2668 & 0.1111 \\ 
\textbf{NEAT} & 0.0335 & 0.0335 & \textbf{0.0835} & \textbf{0.0619} & \textbf{0.1273} & \textbf{0.0798} & 0.2075 & 0.1054 & 0.3403 & 0.1388 \\ 
\textbf{NEAT}+bpr & \textbf{0.0341} & \textbf{0.0341} & 0.0823 & 0.0613 & 0.1227 & 0.0778 & \textbf{0.2163} & \textbf{0.1078} & \textbf{0.3424} & \textbf{0.1395} \\ 
\hline
    \end{tabular}
\end{table*}

\noindent\textbf{Results}: 
We summarize the results of HR@$K$ and NDCG@$K$ for INS dataset (in Tables \ref{tab:metric_1}-\ref{tab:metric_3} ) and WMT dataset (in Tables \ref{tab:metric_4}-\ref{tab:metric_6}).
The best performance for each metric is highlighted in bold.
\textbf{Pop} shows zero HR@$K$ and NDCG@$K$ when $K$ is small.
As aforementioned, popular items are involved in many co-purchase records which are not motivated by complementary relationships.
After removing irrelevant co-purchase records from the dataset by the trustworthy label generation, \textbf{Pop} is less likely to hit a complementary co-purchase.
\textbf{Popco} still achieves reasonable performance on all metrics because it captures the noisy item-to-item complementary relationship via ranking the co-purchased items by their co-purchase frequency with the query item.
\textbf{Item2Vec} and \textbf{Triple2Vec} outperform the frequency-based baselines due to the advantage of item vector representation.
Our model further improves the performance on both HR and NDCG compared with frequency-based baselines and the vector-based baselines.
The results indicate the advantage of modeling the label noise in the co-purchase distribution.

\begin{table*}[t!]
    \centering
    \caption{WMT Labels, p-value = 0.05}
    \label{tab:metric_4}
    \begin{tabular}{l|cc|cc|cc|cc|cc }
    \hline
& \textbf{HR@1} & \textbf{NDCG@1} & \textbf{HR@3} & \textbf{NDCG@3} & \textbf{HR@5} & \textbf{NDCG@5} & \textbf{HR@10} & \textbf{NDCG@10} & \textbf{HR@20} & \textbf{NDCG@20} \\ 
\hline
\textbf{Pop}  & 0.0000 & 0.0000 & 0.0000 & 0.0000 & 0.0000 & 0.0000 & 0.0002 & 0.0001 & 0.0014 & 0.0004 \\ 
\textbf{PopCo} & 0.0069 & 0.0069 & 0.0207 & 0.0148 & 0.0310 & 0.0190 & 0.0506 & 0.0253 & 0.0803 & 0.0328 \\ 
\hdashline
\textbf{CF} & 0.0033 & 0.0033 & 0.0076 & 0.0058 & 0.0105 & 0.0070 & 0.0193 & 0.0098 & 0.0451 & 0.0162 \\ 
\textbf{BPRMF} & 0.0042 & 0.0042 & 0.0108 & 0.0080 & 0.0164 & 0.0103 & 0.0276 & 0.0139 & 0.0505 & 0.0196 \\ 
\textbf{Item2Vec} & 0.0082 & 0.0082 & 0.0200 & 0.0149 & 0.0298 & 0.0189 & 0.0504 & 0.0256 & 0.0818 & 0.0335 \\ 
\textbf{Triple2Vec} & 0.0087 & 0.0087 & 0.0210 & 0.0158 & 0.0294 & 0.0192 & 0.0438 & 0.0239 & 0.0615 & 0.0283 \\ 
\textbf{NEAT} & 0.0120 & 0.0120 & 0.0292 & 0.0219 & 0.0437 & 0.0278 & 0.0715 & 0.0367 & 0.1065 & 0.0455 \\ 
\textbf{NEAT}+bpr & \textbf{0.0121} & \textbf{0.0121} & \textbf{0.0298} & \textbf{0.0221} & \textbf{0.0437} & \textbf{0.0278} & \textbf{0.0717} & \textbf{0.0368} & \textbf{0.1074} & \textbf{0.0459} \\ 
\hline
    \end{tabular}
\end{table*}

\begin{table*}[t!]
    \centering
    \caption{WMT Labels, p-value = 0.01}
    \label{tab:metric_5}
    \begin{tabular}{l|cc|cc|cc|cc|cc}
    \hline
 & \textbf{HR@1} & \textbf{NDCG@1} & \textbf{HR@3} & \textbf{NDCG@3} & \textbf{HR@5} & \textbf{NDCG@5} & \textbf{HR@10} & \textbf{NDCG@10} & \textbf{HR@20} & \textbf{NDCG@20} \\ 
 \hline
\textbf{Pop}  & 0.0000 & 0.0000 & 0.0000 & 0.0000 & 0.0000 & 0.0000 & 0.0001 & 0.0000 & 0.0013 & 0.0003 \\ 
\textbf{PopCo} & 0.0099 & 0.0099 & 0.0291 & 0.0208 & 0.0432 & 0.0266 & 0.0695 & 0.0351 & 0.1080 & 0.0448 \\ 
\hdashline
\textbf{CF} & 0.0042 & 0.0042 & 0.0098 & 0.0074 & 0.0134 & 0.0089 & 0.0244 & 0.0124 & 0.0574 & 0.0206 \\ 
\textbf{BPRMF} & 0.0055 & 0.0055 & 0.0141 & 0.0104 & 0.0212 & 0.0133 & 0.0359 & 0.0180 & 0.0648 & 0.0252 \\ 
\textbf{Item2Vec} & 0.0110 & 0.0110 & 0.0261 & 0.0196 & 0.0388 & 0.0248 & 0.0649 & 0.0332 & 0.1059 & 0.0435 \\ 
\textbf{Triple2Vec} & 0.0117 & 0.0117 & 0.0273 & 0.0207 & 0.0379 & 0.0250 & 0.0563 & 0.0310 & 0.0786 & 0.0366 \\ 
\textbf{NEAT} & 0.0165 & 0.0165 & 0.0393 & 0.0295 & \textbf{0.0583} & \textbf{0.0373} & \textbf{0.0945} & \textbf{0.0490} & 0.1393 & 0.0603 \\ 
\textbf{NEAT}+bpr & \textbf{0.0165} & \textbf{0.0165} & \textbf{0.0401} & \textbf{0.0299} & 0.0582 & 0.0373 & 0.0944 & 0.0490 & \textbf{0.1402} & \textbf{0.0606} \\ 
\hline
    \end{tabular}
\end{table*}

\begin{table*}[t!]
    \centering
    \caption{WMT Labels, p-value = 0.001}
    \label{tab:metric_6}
    \begin{tabular}{l|cc|cc|cc|cc|cc}
    \hline
& \textbf{HR@1} & \textbf{NDCG@1} & \textbf{HR@3} & \textbf{NDCG@3} & \textbf{HR@5} & \textbf{NDCG@5} & \textbf{HR@10} & \textbf{NDCG@10} & \textbf{HR@20} & \textbf{NDCG@20} \\ 
\hline
\textbf{Pop}  & 0.0000 & 0.0000 & 0.0000 & 0.0000 & 0.0000 & 0.0000 & 0.0001 & 0.0000 & 0.0012 & 0.0003 \\ 
\textbf{PopCo} & 0.0140 & 0.0140 & 0.0407 & 0.0293 & 0.0599 & 0.0371 & 0.0948 & 0.0484 & 0.1437 & 0.0607 \\ 
\hdashline
\textbf{CF} & 0.0055 & 0.0055 & 0.0129 & 0.0097 & 0.0173 & 0.0115 & 0.0313 & 0.0160 & 0.0730 & 0.0263 \\ 
\textbf{BPRMF} & 0.0069 & 0.0069 & 0.0177 & 0.0130 & 0.0271 & 0.0168 & 0.0461 & 0.0229 & 0.0831 & 0.0322 \\ 
\textbf{Item2Vec} & 0.0145 & 0.0145 & 0.0342 & 0.0257 & 0.0504 & 0.0324 & 0.0837 & 0.0431 & 0.1365 & 0.0563 \\ 
\textbf{Triple2Vec} & 0.0156 & 0.0156 & 0.0356 & 0.0271 & 0.0491 & 0.0327 & 0.0725 & 0.0402 & 0.1010 & 0.0474 \\ 
\textbf{NEAT} & \textbf{0.0226} & \textbf{0.0226} & 0.0524 & 0.0396 & \textbf{0.0771} & \textbf{0.0498} & 0.1237 & 0.0648 & 0.1806 & 0.0792 \\ 
\textbf{NEAT}+bpr & 0.0223 & 0.0223 & \textbf{0.0532} & \textbf{0.0399} & 0.0771 & 0.0497 & \textbf{0.1241} & \textbf{0.0649} & \textbf{0.1815} & \textbf{0.0794} \\ 
\hline
    \end{tabular}
\end{table*}

\subsection{Ablation Study of \NEATe}
Our model can be extended with user embeddings to model the complementary relationship from the user-item-level co-purchase data.
To study the extensibility of our model and the influence of involving user embeddings, we compute HR@$K$ and NDCG@$K$ for \NEAT and \NEATe+bpr, $K = \{1, 3, 5, 10, 20\}$.
The results are summarized in Tables \ref{tab:metric_1}-\ref{tab:metric_6}.
We can see that both \NEAT and \NEATe+bpr perform similarly but \NEATe+bpr outperforms \NEAT in most cases when: 
(1) $K$ becomes larger or (2) number of items increases from INS dataset to WMT dataset.
This indicates that including user-item-level signals improves the model performance especially when the number of items is large.

\begin{figure}
    \centering
    \includegraphics[width=\linewidth]{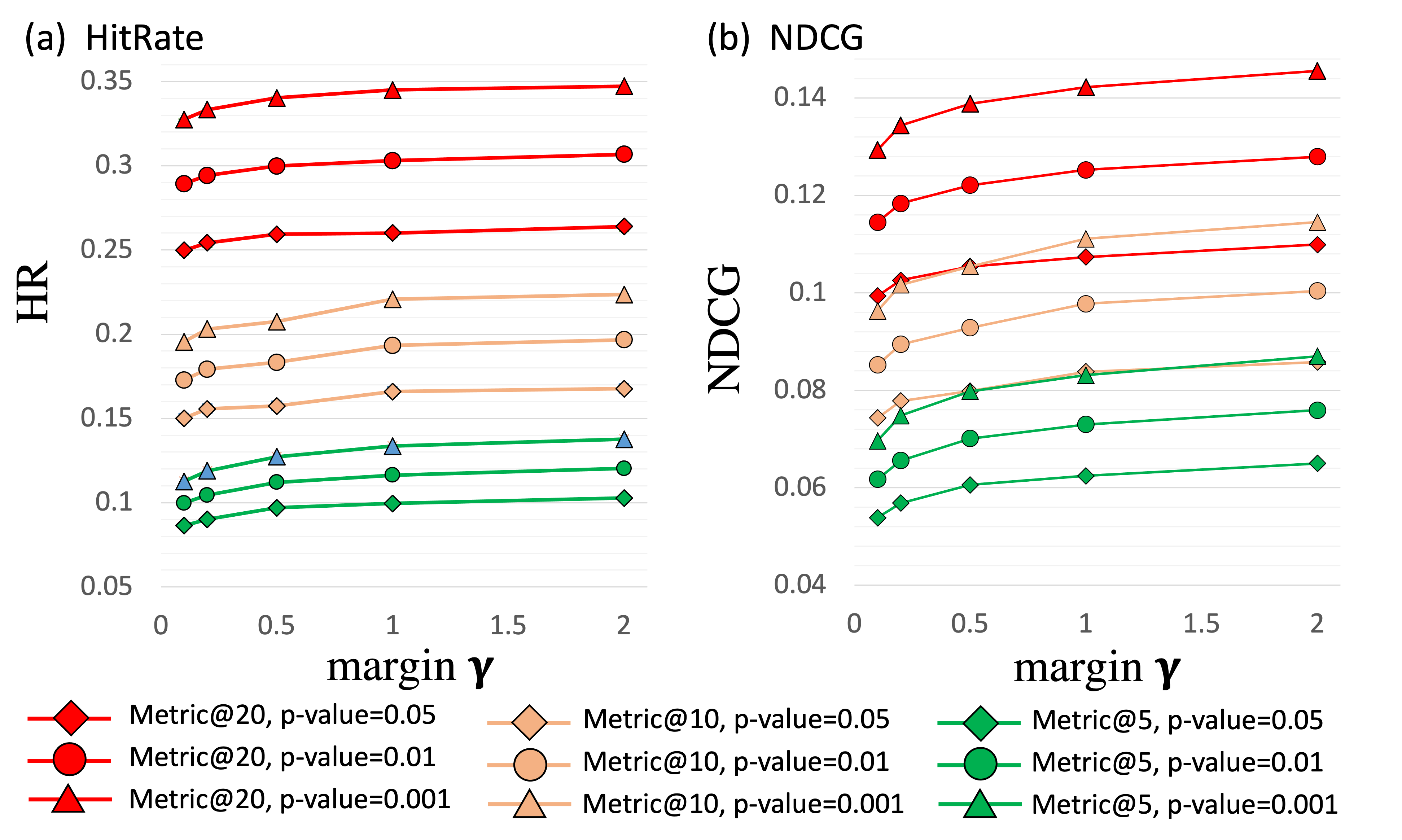}
    \caption{Analysis of the margin $\gamma$ on three label sets, p-value = $\{0.05, 0.01, 0.001\}$, for metric@$\{5, 10, 20\}$ of Hit-Rate and NDCG on INS dataset.}
    \label{fig:gamma_analysis1}
\end{figure}

\begin{figure}
    \centering
    \includegraphics[width=\linewidth]{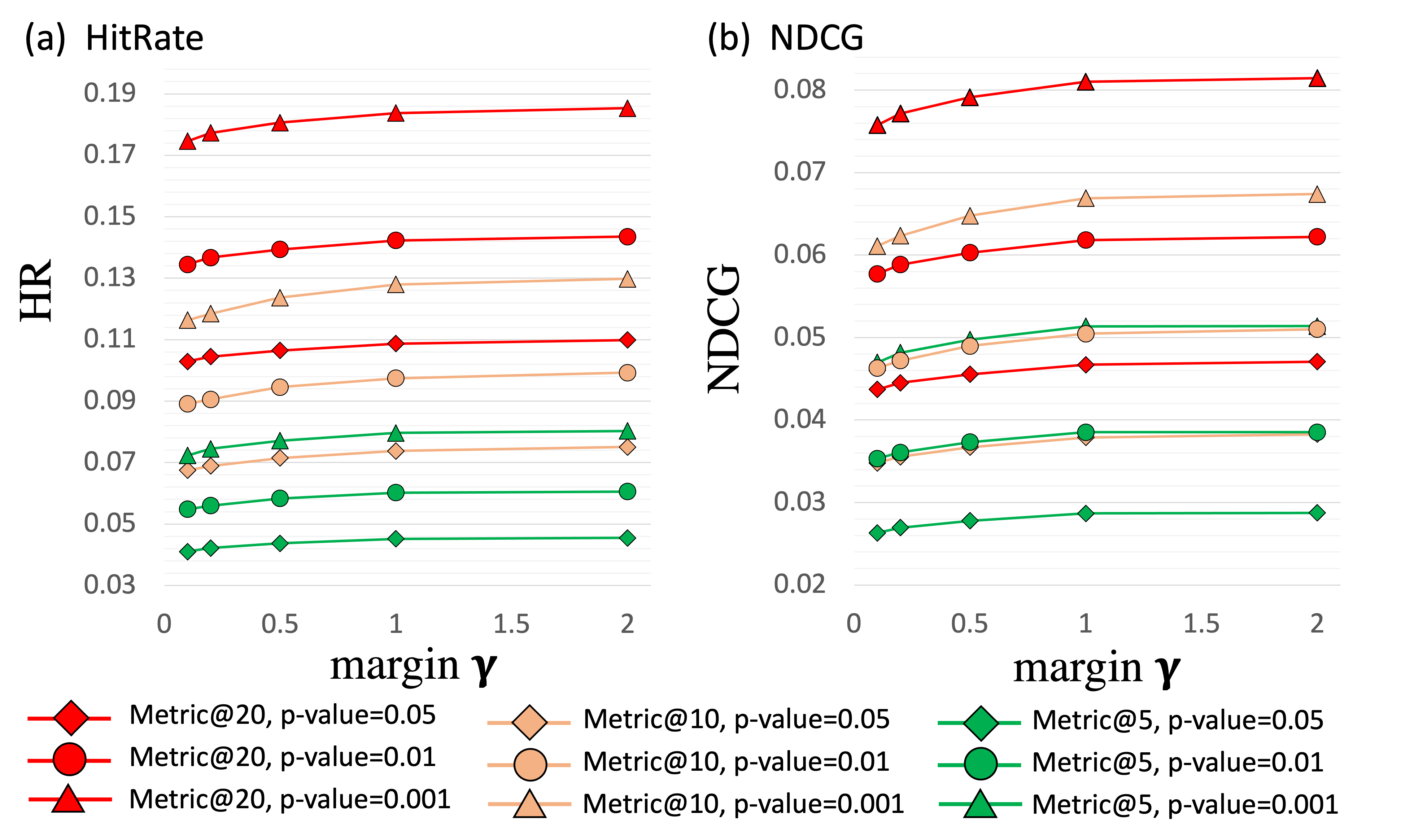}
    \caption{Analysis of the margin $\gamma$ on three label sets, p-value = $\{0.05, 0.01, 0.001\}$, for metric@$\{5, 10, 20\}$ of Hit-Rate and NDCG on WMT dataset.}
    \label{fig:gamma_analysis2}
\end{figure}

\subsection{Sensitivity Analysis of the Margin $\gamma$}
We conduct experiments on \NEAT with different margins $\gamma = \{0.1, 0.2, 0.5, 1.0, 2.0\}$ on the three label sets of INS dataset and WMT dataset respectively.
We report HR@$K$ and NDCG@$K$ for evaluation with $K = \{5, 10, 20\}$ and summarize the results in Figure \ref{fig:gamma_analysis1} and \ref{fig:gamma_analysis2}. The results indicate that the model is in favor of a larger margin. 


\subsection{Case Study: Item Representation as a Distribution}
To test whether the Gaussian embedding of items could capture the variation of items in their co-purchase, we focus on three items, \texttt{Whole Milk}, \texttt{Cereal} and \texttt{Organic Tortilla Chips} in INS dataset, and study the relationship between item Gaussian embeddings and complementary relationships when \texttt{Whole Milk} becomes the query item.
On one hand, the cosine similarity of $\mbox{\boldmath${\mu}$}_{\text{\texttt{Whole Milk}}}$ and $\mbox{\boldmath${\mu}$}_{\text{\texttt{Cereal}}}$ is larger then that of $\mbox{\boldmath${\mu}$}_{\text{\texttt{Whole Milk}}}$ and $\mbox{\boldmath${\mu}$}_{\text{\texttt{Organic Tortilla Chips}}}$, which aligns with the expectation of stronger complementary relationship between \texttt{Whole Milk} and \texttt{Cereal}.
On the other hand, the query item \texttt{Whole Milk} which has higher popularity than \texttt{Cereal} and \texttt{Organic Tortilla Chips} in INS dataset also shows higher variation (indicated by the determinant of the spherical covariance matrix). 
In particular, the $\det(\mbox{\boldmath$\Sigma$}_{\text{\texttt{Whole Milk}}})$ is 30 times larger than $\det(\mbox{\boldmath$\Sigma$}_{\text{\texttt{Cereal}}})$  and is 547 times larger than $\det(\mbox{\boldmath$\Sigma$}_{\text{\texttt{Organic Tortilla Chips}}})$.
This also aligns with our expectation of their variation since \texttt{Whole milk} (35633 purchases) is more popular than \texttt{Cereal} (12184 purchases) and \texttt{Organic Tortilla Chips} (13776 purchases) in INS dataset and hence more likely to form irrelevant co-purchases.
